\documentclass[default,iicol]{sn-jnl}
\usepackage[version=4]{ mhchem }
\usepackage{graphicx}%
\usepackage{siunitx}
\usepackage{multirow}%
\usepackage{amsmath,amssymb,amsfonts}%
\usepackage{amsthm}%
\usepackage{mathrsfs}%
\usepackage[title]{appendix}%
\usepackage{xcolor}%
\usepackage{textcomp}%
\usepackage{manyfoot}%
\usepackage{booktabs}%
\usepackage{algorithm}%
\usepackage{algorithmicx}%
\usepackage{algpseudocode}%
\usepackage{listings}%
\usepackage{mhchem} 
\usepackage{lineno}
\usepackage{makecell}
\usepackage{float}
\usepackage[numbers]{natbib}
\theoremstyle{thmstyleone}%

\theoremstyle{thmstyletwo}%

\theoremstyle{thmstylethree}%

\begin{document}

\title[Article Title]{Efficient transmutation of long-lived fission products in a Gamma Factory beam driven advanced nuclear energy system}
\author[1,2]{\fnm{Baolong} \sur{Hu}}
\author*[3,4]{\fnm{Mieczyslaw Witold} 
\sur{Krasny}}\email{mieczyslaw.witold.krasny@cern.ch}
\author[5]{\fnm{Wies{\l}aw} \sur{P{\l}aczek}}
\author[1,2]{\fnm{Yun} \sur{Yuan}}
\author[1,2]{\fnm{Xiaoming} \sur{Shi}}
\author[1,2]{\fnm{Kaijun} \sur{Luo}}
\author*[1,2]{\fnm{Wen} \sur{Luo}}\email{wenluo-ok@163.com}

\affil*[1]{\orgdiv{School of Nuclear Science and Technology}, \orgname{University of South China}, \orgaddress{\postcode{421001}, \city{Hengyang}, \country{China}}}

\affil*[2]{\orgdiv{Key Laboratory of Advanced Nuclear Energy Design and Safety}, \orgname{Ministry of Education}, \orgaddress{\postcode{421001}, \city{Hengyang}, \country{China}}}

\affil*[3]{\orgdiv{LPNHE}, \orgname{University Paris Sorbonne}, \orgaddress{\city{Paris}, \postcode{75005}, \country{France}}}

\affil*[4]{\orgdiv{CERN},  \orgaddress{\city{Geneva}, \postcode{1211}, \country{Switzerland}}}

\affil[5]{\orgdiv{Faculty of Physics, Astronomy and Applied Computer Science and Mark Kac Center for Complex Systems Research}, \orgname{Jagiellonian University}, \orgaddress{ul. {\L}ojasiewicza 11, \postcode{30–348} \city{Krakow},  \country{Poland}}}


\abstract{
The Gamma Factory (GF) project aims to generate high-intensity $\gamma$-ray beams of tunable energy and relatively small energy spread. Such beams can be optimized to generate an intense photo-neutron source, capable of driving an advanced nuclear energy system (ANES) for nuclear waste transmutation and supplying electrical power that is necessary for the GF operation mode of the Large Hadron Collider storage ring. In this study, we investigate the feasibility of driving ANES with the GF beam
which is optimized to maximize the neutron production rate. The dependence of the ANES thermal power on the distance between the positions of the ANES and the GF $\gamma$-ray source is evaluated. For the $\gamma$-ray beam reaching the intensity of $\sim$$10^{19}$ photons per second, the ANES thermal power could exceed $500\,$MWt. Under the assumption that ANES operates over $20$ years, the transmutation rate could reach $30\%$ for five typical long-lived fission products (LLFPs): $^{79}$Se, $^{99}$Tc, $^{107}$Pd, $^{129}$I, $^{137}$Cs. Our comparative studies show that although the neutron production efficiency of the GF $\gamma$-ray beam (per MW of the beam power) is approximately $14$ times lower than that of the  $500\,$MeV proton beam, the overall net ANES power production efficiency for the GF beam driver scheme could be comparable to that of the proton beam driver scheme,  while providing additional transmutation capacity, not available for the proton beam driven scheme. It is suggested that the GF-based ANES could provide a viable solution for the efficient transmutation of LLFPs without isotopic separation.
}


\maketitle 

\newpage

\section{Introduction}



Nuclear power is considered as a strategic choice to solve the future energy supply and to ensure sustainable economic and social development \cite{N07wang2018overview,N08chen2018prospects}. 
At present, it supplies around one-quarter of the world's clean electricity \cite{N06PerformanceReport2023}. Along with the development of nuclear power, the generation of spent nuclear fuel (SNF) also grows accordingly. While less than one third of the SNF inventory is reprocessed each year \cite{N09fukuda2003iaea}, about $10000\,$t remains to be disposed \cite{N10sworth2023spent}. How to safely dispose the stored and newly generated SNF is becoming an urgent issue to ensure the sustainable development of nuclear power \cite{N11gu2021latest}. 

To solve the SNF problem, the “partitioning-transmutation” concept has been suggested in the 1990s \cite{N12experimental1995,N13rubbia1995conceptual,N14salvatores2011radioactive,N15OECD,N16zhan2012advanced}. After recovering U and Pu from SNF by PUREX process, high-level radioactive hazards are mainly constituted of radiotoxic transuranics (TRUs) and long-lived fission products (LLFPs), including $^{79}$Se, $^{93}$Zr, $^{99}$Tc, $^{107}$Pd, $^{129}$I, $^{135}$Cs, $^{137}$Cs. 

Transmutation of LLFPs mainly relies on neutron capture reactions or photo-neutron reactions \cite{N18wang2016photo,N17wang2017transmutation,N19optimization2017,N20rehman2018comparison}. LLFPs can be transmuted into short-lived or stable nuclei by using a high-flux neutron or photon source. To perform an efficient transmutation of LLFPs, isotopic separation becomes a necessity since LLFPs have relatively small isotopic abundances. However, no isotope-separation system for high-level nuclear wastes is so far technologically and economically feasible on an industrial scale \cite{N21imasaki2008gamma}. 

Recently, a novel concept of ANES has been proposed for transmuting LLFPs efficiently without isotopic separation \cite{N22sun2022transmutation}. Such ANES consists of a photo-neutron source (PNS) and a subcritical reactor. 
Seven LLFPs are loaded in the reactor core for the neutron-driven transmutation. 

For the ANES operating mode at a thermal power of $500\,$MWt, its PNS must be driven
by a high-intensity photon beam, of 10$^{19}$ photons per second. Such a flux is many orders of magnitudes higher than that of the existing and the future electron-beam-driven laser-Compton scattering sources \cite{N01Proposal2015,N04budker2022expanding}. 

The Gamma Factory (GF)  proposal \cite{N01Proposal2015} is currently being studied within the CERN Physics Beyond Colliders (PBC) framework \cite{N02jaeckel2020quest,N03krasny2018cern}. PBC explores  complementary options for the future CERN particle-beam-driven research programme. One of the multiple goals of GF is to generate high-intensity $\gamma$-ray beams of tunable energy and relatively small energy spread, with energies up to $\approx 400\,$MeV,  and photon fluxes exceeding those of the currently available $\gamma$ sources by many orders of magnitudes \cite{N01Proposal2015}. The leap in the GF photon flux can be achieved by colliding laser photons with the atomic beams of partially stripped ions circulating in one of the CERN storage rings. 

The GF photon beam can be optimized to maximize the production rate of $7\,$--$\,20\,$MeV photons. Photons in this energy range can be produced by atomic beams stored in the Large Hadron Collider (LHC) rings. 
They can excite the giant dipole resonance (GDR) in medium-mass and heavy nuclei \cite{N05berman1975measurements}. 
Since the cross section for these processes is in the barn range, a large fraction of the LHC radio-frequency (RF) power can be converted into the power of neutron and radioactive-ion beams by colliding the GF photon beam with stationary targets \cite{N01Proposal2015,N04budker2022expanding,Nichita:2021iwa}. 

The GF $\gamma$-ray beam can open new avenues: (1) for the development of the photon-beam-driven ANES, and (2) for the highly efficient transmutation of nuclear waste \cite{Krasny:2023ptc}.

\begin{figure*}[htbp]
\centering
\includegraphics[width=0.8\hsize]{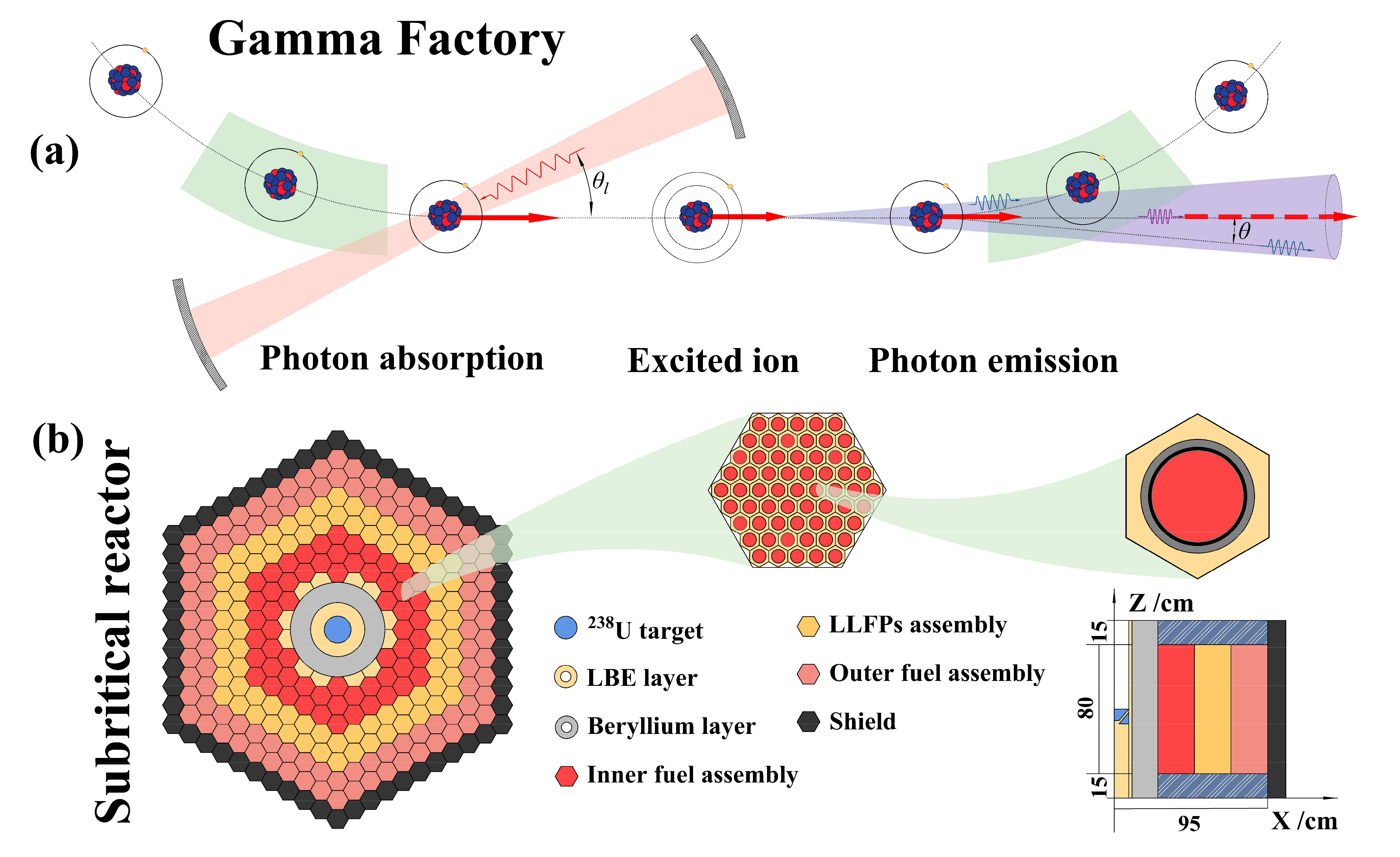}
\caption{The concept of the GF-driven ANES: (a) the resonant generation of the photon beam in GF and (b) the front and side views of the subcritical reactor layout.}
\label{fig1}
\end{figure*}

In this paper, we present our exploratory studies of the GF-based ANES, schematically shown in Fig. \ref{fig1}.  The GF $\gamma$-ray beam produces an intense PNS. Such PNS then delivers neutrons to the subcritical core of the ANES,  enabling both the fission energy generation and the nuclear transmutation. Since the operation of GF requires a stable electrical supply of approximately $50\,$MW, it is assumed in the following that the GF-based ANES produces comparable or higher electric output power to fulfill such a demand. This condition specifies the required minimal power of the GF driver beam.

The paper is organised as follows. An optimized neutron generation scheme is first introduced with the aim of reducing the $\gamma$-ray intensity requirement. The dependence of the ANES thermal power on the distance $d$ between the positions of the ANES and GF $\gamma$-ray source is then evaluated. Under the assumption that the ANES operates over $20$ years, the transmutation rates for seven selected LLFPs are evaluated. Finally, the respective advantages of the photon-beam-driven and the proton-beam driven ANES are discussed. Some supplementary information on the used computational methods and the studied materials is provided in the appendices.

\section{Gamma Factory $\gamma$-ray beam}
\label{sec:GF}

The  GF idea is to use bunches   
of partially stripped ion (PSI) beam, circulating  in the LHC, and to collide them with laser photon pulses stored in a Fabry-Perot (FP) cavity; cf.\ Fig.\ref{fig1}(a).  In the process of a resonant absorption of a laser photon by PSI, followed by a spontaneous atomic-transition emission of a secondary photon, the initial laser-photon frequency is boosted by a factor of up to $4\gamma^{2}_{L}$, where $\gamma_{L}$ is the relativistic Lorenz factor of the partially stripped ion beam \cite{N01Proposal2015}. The GF $\gamma$-ray beams produced by the atomic-transition emissions can push the intensity limits of the presently operating light sources by at least $7$ orders of magnitude in the particularly interesting $\gamma$-ray energy domain of $1\, \mathrm{MeV}\lesssim E \lesssim 400\,$MeV. The GF $\gamma$-ray beam is characterized by a strong correlation between the photon energy $E$ and its emission angle $\theta$. The energy tunability of the GF $\gamma$-ray beam is achieved by simultaneously tuning the Lorenz factor $\gamma_{L}$ of the PSI beam and by adjusting the wavelength $\lambda$,  or the incident angle $\theta_{l}$,  of the stored laser photons to assure the resonant photon absorption.  The $\gamma$-ray energy can be expressed as \cite{Wen2016}:
\begin{equation}
E(\theta )=\cfrac{E_{\rm max}}{2{\gamma_L^{2}}(1-\beta \cos\theta)}\,,  
\label{eq:Etheta1}
\end{equation}
where $E_{\rm max} = (1+\beta)\gamma_L\hbar\omega^{\prime} $, $\beta$ is the ion speed normalized by the speed of light in the vacuum, and $\hbar\omega' = \hbar \omega \gamma_L (1 + \beta\cos\theta_l)$, where $\hbar \omega$ is the laser photon energy. 
For small values of $\theta$ and $\theta_{l}$, it can be approximated as:
\begin{equation}
E(\theta )\approx \cfrac{E_{\rm max}}{1+\gamma_L^{2}\theta^{2}}\approx \cfrac{4\gamma_L^{2}\hbar\omega}{1+\gamma_L^{2}\theta^{2}}\,.
\label{eq:Etheta2}
\end{equation}

Up to now, all the high-intensity $\gamma$-ray beams have been generated by colliding laser photons with electron beams. The cross section for the inverse Compton-scattering process is by a factor of up to 10$^{9}$ smaller than the resonant atomic photo-excitation cross section of PSIs. This is the main reason why the intensity of the PSI-beam-driven $\gamma$-ray beam can be higher than that of the electron-beam-driven ones by many orders of magnitude \cite{N01Proposal2015}. 
 
The optimal PSIs for the ANES application are the helium-like calcium ions, Ca$^{18+}$. A detailed discussion of the production and storage of the Ca beam in the CERN accelerator complex can be found in \cite{Krasny:2020wgx}. A specific choice of the atomic transition:  $1s^2\,{}^1\mathrm{S}_0 \rightarrow 1s2p\,{}^1\mathrm{P}_1$,  selected for the studies presented in this paper,  maximizes the fraction of the GF photon flux in the relevant GDR energy region of $7\,$--$\,20\,$MeV.

In all the published to date GF applications studies (see e.g.\ \cite{Krasny:2023ptc} and references quoted therein), the required photon flux has been limited to $\sim\,$$10^{17}\,\gamma/$s. Such intensities can be reached within the current CERN accelerator infrastructure operation constraints. The GF-driven ANES case, as we shall discuss in this paper,  requires increasing photon flux by two orders of magnitude. To achieve such a goal, the following two upgrades to the LHC RF system will have to be made:
\begin{itemize}
    \item Currently, the LHC RF power is generated by  $300\,$kW/$400\,$MHz klystrons supplying $4.8\,$MW power  
to the LHC beams \cite{Bruning:2004ej,ZurbanoFernandez:2020cco}.  
    In order to continuously produce  $\sim\,$$10^{19}\,\gamma/$s by the stored Ca$^{+18}$ beam,  the klystron power will  have to be increased by a factor of 4%
    \footnote{LEP2, an earlier storage ring installed in the LHC tunnel, was already operating with 
    $44$ klystrons, each of $1.3\,$MW power, delivering the $57\,$MW RF-power  
    to the electron beam.} 
    to fully compensate for the ion energy loss in the process of $\gamma$-ray emission.
    \item The present LHC circumferential voltage of $8$ cavities (two cryomodules) is $C_v =  16\,$MV \cite{ZurbanoFernandez:2020cco}. For the $\sim\,$$10^{19}\, \gamma/$s beam produced by He-like Ca ion bunches, $C_v$ will  have to be increased to the value of $\sim\,$$180\,$MV%
    \footnote{LEP2 was operating with  $C_v =  3650\,$MV to compensate for the average  $\sim\,$$3\,$GeV per-turn energy loss of each of the electrons/positrons.}.
\end{itemize}

For the current laser-technology accessible case of the $500\,$ps long,  $5\,$mJ pulse-power and $20\,$MHz repetition-rate “Erbium laser” photon pulses,  stored in the FP cavity and colliding with the Ca$^{18+}$ ion bunches of $10^9\,$ions/bunch at the zero-degree crossing geometry, the maximal achievable $\gamma$-ray production rate could reach the value of $\sim\,$$10^{18}\,\gamma/$s,  which is a factor of $10$ lower than that required for the ANES application. The proposed solution to reach the requisite $\gamma$-beam intensity in collisions of laser pulses with PSIs at a realistic crossing angle of $1\,$deg requires the implementation of $20$ identical FP cavities, similar to those designed for the GF Proof-of-Principle experiment \cite{GF-PoP-LoI:2019}. In the studies presented in this paper, these cavities are assumed to be placed in one of the LHC straight sections. The large,  $\beta^* = 50\,$m,  collision optics would be required  to minimize the variation of the transverse ion-bunch size in the accelerator straight section where the FP cavities are implemented. 

The full set of parameters used in the GF $\gamma$-ray beam simulations,
including the specification of the Ca$^{18+}$-beam parameters, the laser-pulse parameters, and the optics of the  interaction point (IP) region 
is presented in Table \ref{tab:YbHepars}. A PSI beam-cooling scheme allowing to reduce the PSI-bunch length to $1.5\,$cm is discussed in \cite{Krasny:2020wgx}.

\begin{figure*}[htbp]
    \centering
    \includegraphics[width=0.8\textwidth]{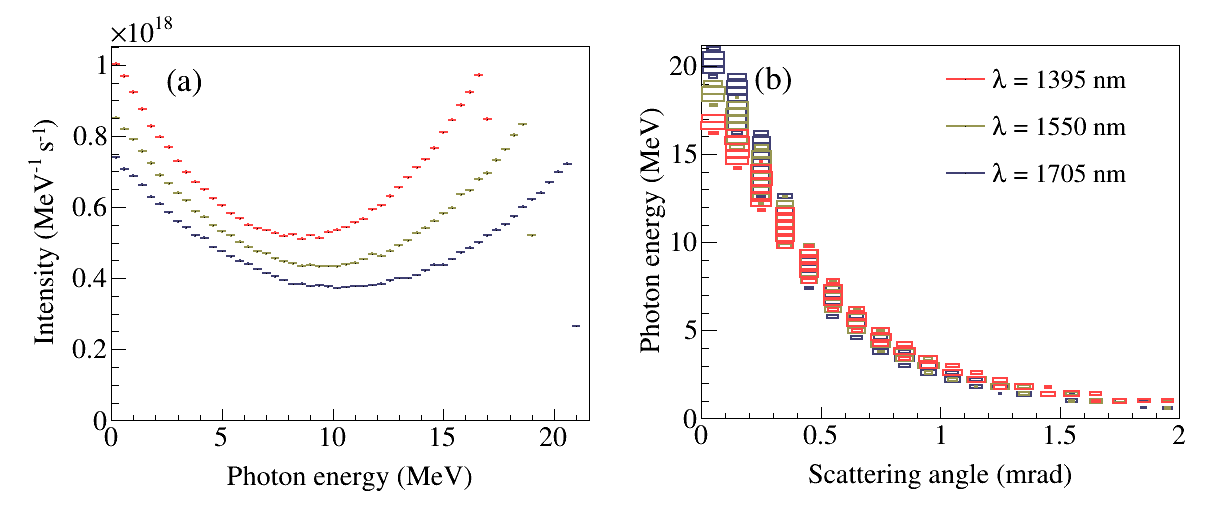}
    \caption{The energy distributions of the GF $\gamma$-ray beams produced at IP for three values of laser wavelength $\lambda$ (a) and the dependence of the GF $\gamma$-ray energy on the scattering angle $\theta$ (b).}
    \label{fig2}
\end{figure*}

As shown in Fig. \ref{fig2}, the integrated power of the GF $\gamma$-ray beam produced in such a configuration is $16.6\,$MW.  Its spectrum covers the required GDR and photo-fission domain of energies and reaches a sufficiently large beam intensity and a sufficiently small angular divergence to be used as the driver of PNS.
 
\begin{table*}[ht]
\centering
\caption{The parameters of the cooling configuration in the SPS for the {\sf GF-CAIN} \cite{GF-CAIN} simulations of 
the He-like Ca beam, the Erbium laser with the photon wavelength $\lambda\,$$=$$\,1550\,$nm 
and the atomic transition $1s^2\,{}^1\mathrm{S}_0 \rightarrow 1s2p\,{}^1\mathrm{P}_1$ for ${}^{40}_{20}\mathrm{Ca}^{18+}$.
}
\small\addtolength{\tabcolsep}{8.pt}
\renewcommand{\arraystretch}{1.1}
        \begin{tabular}{lr}\hline
                PSI beam  & ${}^{40}_{20}\mathrm{Ca}^{18+}$\\
                \hline
                $m$ -- ion mass & $37.332\,$GeV/c$^2$ \\
                $E$ -- mean energy & $91.07$ TeV \\
                $\gamma_L=E/mc^2$-- mean Lorentz relativistic factor & $2439.5$ \\
                $N$ -- number ions per bunch & $3\times 10^9$ \\
                $\sigma_E/E$ -- RMS relative energy spread & $2\times 10^{-4}$  \\
                $\beta_x=\beta_y$ -- $\beta$-function at interaction-point (IP)& $50\,$m\\
                $\varepsilon_x=\varepsilon_y$ --  beam geometric emittance & $3\times 10^{-10}\,\mathrm{m \cdot rad}$\\
                $\sigma_z$ -- RMS bunch length & $1.5\,$cm\\
                Bunch repetition rate & $20\,$MHz\\
                \hline
                Laser & Erbium\\
                \hline
                $\lambda$ -- photon wavelength & $1550\,$nm  \\
                $\hbar\omega$ -- photon energy & $0.799898\,$eV \\
                $\sigma_{\lambda}/\lambda$ -- RMS relative band spread & $2\times 10^{-4}$  \\
                $U$ -- single pulse energy at IP & $5\,$mJ \\
                $\sigma_x=\sigma_y$ -- RMS transverse intensity distribution at IP & $300\,\mathrm{\mu m}$\\
                $\sigma_z$ -- RMS pulse length & $2.99792\,$cm \\
                $\theta_l$ -- collision angle & $1\deg$ \\
                $N_l$ -- number of laser stations  & $20$  \\
                \hline
                 Atomic transition of ${}^{40}_{20}\mathrm{Ca}^{18+}$ & $1s^2\,{}^1\mathrm{S}_0 \rightarrow 1s2p\,{}^1\mathrm{P}_1$\\
                \hline
                $\hbar\omega'_r $ -- resonance energy & $3.902\,$keV\\
                $\tau'$ -- mean lifetime of spontaneous emission & $6\times 10^{-15}\,$s\\
                $g_1, g_2$ -- degeneracy factors of the ground and excited states& $1,3$\\
               $d\sigma/d\Omega_{\gamma}^{'}$ -- angular distribution of emitted photons in ion rest-frame &
               $\propto \left(1 + \cos^2\theta_{\gamma}^{'}\right)$\\
                $\hbar\omega_{1}^{\rm max}$ -- maximum emitted photon energy & $19.04\,$MeV\\
                \hline
        \end{tabular}
\label{tab:YbHepars}
\end{table*}

\begin{figure*}[htbp]
\centering
    \includegraphics[width=0.9\textwidth]{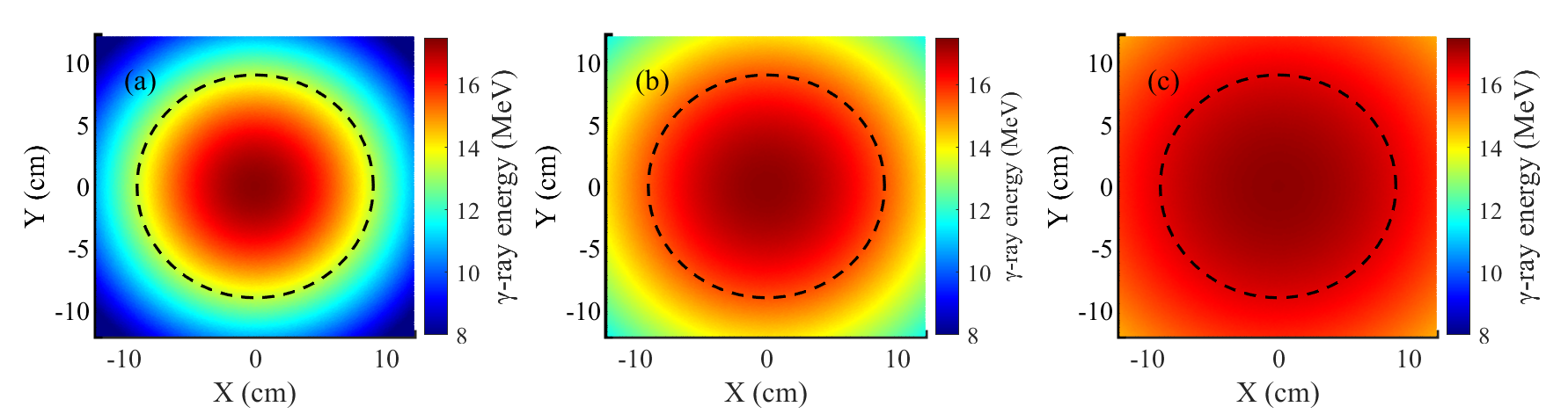}
    \caption{The spatial distributions of the GF $\gamma$-ray beam diagnosed at $300$, $500$ and $800\,$m away from IP, respectively. The dashed circle indicates the position of the $^{238}$U target.}
    \label{fig3}
\end{figure*}

The transverse-plane distributions of the GF $\gamma$-ray beam hits at the three studied distances the $^{238}$U target which produces the required PNS and has a fixed  radius of $9\,$cm, from the $\gamma$-beam production zone of $d = 300$, $500$ and $800\,$m are shown in Fig.~\ref{fig3}. 
Since the LHC superconducting magnets have to be shielded from the excessive neutron fluxes, the ANES system cannot be placed at distances smaller than $300\,$m from the GF $\gamma$-ray source production zone.  
 
Fig.~\ref{fig3} demonstrates the effects reflecting the divergence of the GF $\gamma$-ray beam and the correlation of the $\gamma$-ray beam energy with its emission angle. As the distance to the ANES system increases, the energies of those $\gamma$-rays that hit the PNS target rise. 

\section{Photo-neutron production}
\label{sec:photo-neutron}

The neutron production rate of PNS per $\gamma$-ray  $P_{\gamma n}$ depends on the $\gamma$-ray spectral distribution and the GDR cross section. The neutrons are produced in both the photo-fission \cite{N27shi2024geant4} and photo-neutron processes. In our studies, the neutron production cross sections $\sigma _{x  n}$ is calculated by the {\sf TALYS} software~\cite{N28Talys2012}. For the $^{238}$U target, 
$\sigma _{x  n}$ reaches its maximal value of $\approx\,$$730\,$mb at the photon energy of $14\,$MeV. For the photon energy ranging from $10.46\,$MeV to $16.39\,$MeV, $\sigma _{x  n}$ exceeds $1/2$  of its maximal value. This energy range is thus optimal for photo-neutron production. 

Fine-tuning of the laser wavelength%
\footnote{The wavelength tunability of the Erbium laser is discussed e.g.\ in \cite{liu2024fully}.}%
and of the corresponding resonant $\gamma_L$ of the Ca$^{+18}$ beam allows to maximize the number of photons in the optimal energy range for PNS placed at various distances $d$ from the GF photon source. 

\begin{figure}[htbp]
\centering
    \includegraphics[width=.5\textwidth]{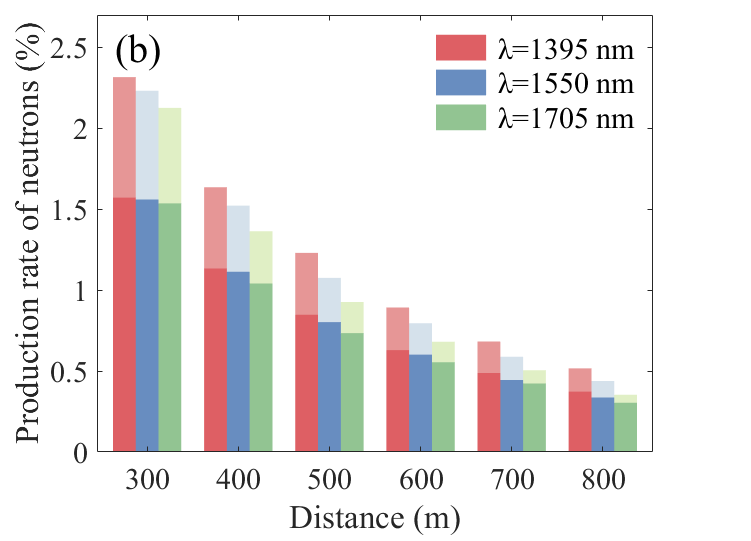}
    \caption{The photo-neutron generation rate $P_{\gamma n}$ as a function of the distance $d$ between the GF $\gamma$-ray source and PNS. The dark and light regions represent the neutron partitions induced by photo-neutron and photo-fission reactions, respectively.}
    \label{fig4}
\end{figure}

The photo-neutron production rate $P_{\gamma n}$ is shown in Fig.~\ref{fig4} as a function of the distance $d$ for three different laser wavelength settings. It is shown that,
for the fixed PNS target radius, $P_{\gamma n}$ decreases with the increasing $d$  due to the $\gamma$-ray beam divergence. 
For $\lambda\,$$=$$\,1395\,$nm, $P_{\gamma n}$ is marginally higher for the other two laser wavelengths.

Our simulations show that approximately 30$\%$ of the total number of photo-neutrons can be attributed to the photo-fission reactions.  $P_{\gamma n}$ reaches the value $0.023$ which by a factor of $2$ higher than that given by X.~Sun\ {\it et al.}~\cite{N22sun2022transmutation}. This is because the $^{238}$U target nucleus, considered in our studies,  has a higher neutron-production cross section as compared to the CsI target used in \cite{N22sun2022transmutation}.
Therefore, for the $^{238}$U target, the $\gamma$-ray intensity that is required to drive ANES operating at the same thermal power could be reduced accordingly.

\section{Performance of ANES}
\label{sec:performance}

\begin{figure}[htbp]
\centering
    \includegraphics[width=.45\textwidth]{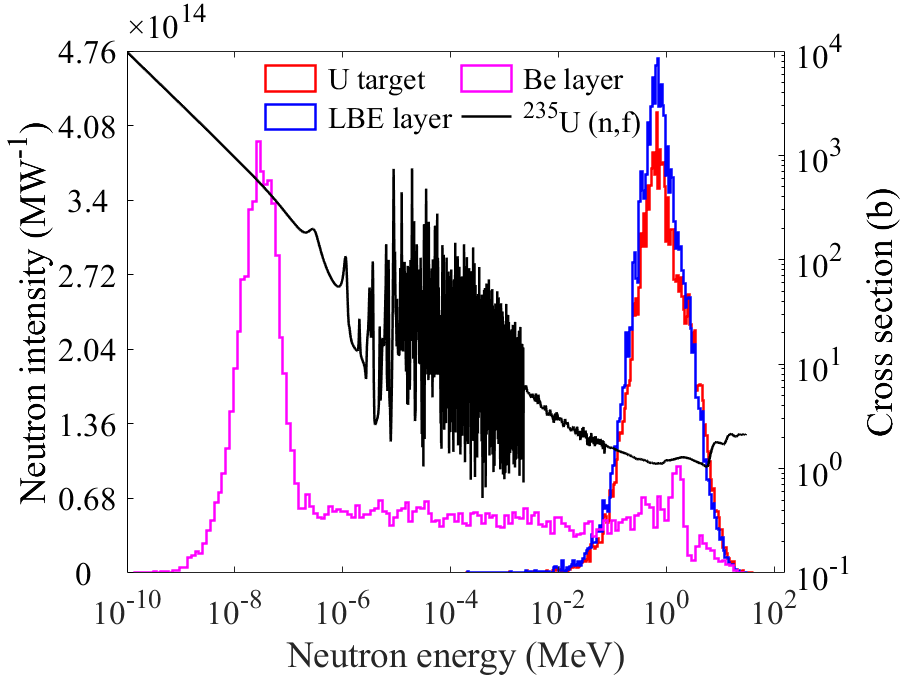}
    \caption{Spectral distributions of neutrons escaped from different layers of ANES for the GF $\gamma$-ray beam with $\lambda\,$$=$$\,1550\,$nm, together with the cross-section curve of the $^{235}$U(n,f) reaction. The thicknesses of the LBE and Be layers used in the {\sf Geant4} simulations are $T_{\rm LBE}\,$$=$$\,2\,$cm and $T_{\rm Be}\,$$=$$\,16\,$cm, respectively.}
    \label{fig5}
\end{figure}

The geometrical layout of ANES, cf.~Fig.~\ref{fig1}(b), is almost the same as the one shown in \cite{N22sun2022transmutation}. The target material used for PNS is changed to $^{238}$U. In addition, the enrichment of $^{235}$U loaded in the fuel assembly is changed to the value of $25.25\%$, which leads to the initial effective neutron multiplication factor $k_{\rm eff}\,$$=$$\,0.979$. PNS produces mainly fast neutrons with energies peaked at a MeV-level. Neutrons escaping from the $^{238}$U target have to be moderated into thermal or epithermal neutrons which can be readily absorbed by the fuel assemblies, inducing the nuclear fission process. A lead-bismuth eutectic (LBE) layer and a beryllium (Be) layer are used for neutron moderation and multiplication, respectively (see Fig.~\ref{fig1}). The former also plays a key role in cooling the $^{238}$U target. The optimised thicknesses for these layers are $T_{\rm Be}\,$$=$$\,16\,$cm and $T_{\rm LBE}\,$$=$$\,2\,$cm. The spectra of neutrons emitted from the PNS $^{238}$U target, the LBE coolant and the Be moderator are presented in Fig.~\ref{fig5}. 

The GF $\gamma$-ray beam, for $\lambda\,$$=$$\,1550\,$nm and $d\,$$=$$\,300\,$m, produces $\sim\,$$1.6$$\times$10$^{16}$ neutrons/s per MW of beam power which escape from the PNS target. 

The cross-section curve of $^{235}$U(n,f) reaction is also shown in Fig.~\ref{fig5}. One can see that the neutron spectrum softens significantly in the moderator. 
The moderated neutrons have more than two orders of magnitudes higher fission,  (n,f),  reaction cross sections than the PNS-originated fast neutrons. 
Their supply increases the neutron flux in the reactor core such that the effective {\it subcritical} neutron multiplication factor \cite{N30xoubi2018neutronic} reaches the maximal value of $k_{s}$ = 0.993
for ANES placed at a distance of $300\,$m from the GF photon source.

The overall efficiency of the GF photon beam in the increasing 
 neutron flux in the reactor core is often quantified in terms 
of the $k_{s}$ and $k_{\rm eff}$ reactor parameters by the {\it neutron worth} \cite{N29fang2021theoretical} parameter defined as: 
\begin{equation}
\varphi = \cfrac{1-1/k_{\rm eff}}{1-1/k_{s}}\,.
\label{eq:varphi}
\end{equation}
For the ANES system discussed in this paper $\varphi \approx$ 3.07. It is two times higher than that given in \cite{N22sun2022transmutation}, for the same integrated photon flux.

\begin{figure}[htbp]
\centering
    \includegraphics[width=.5\textwidth]{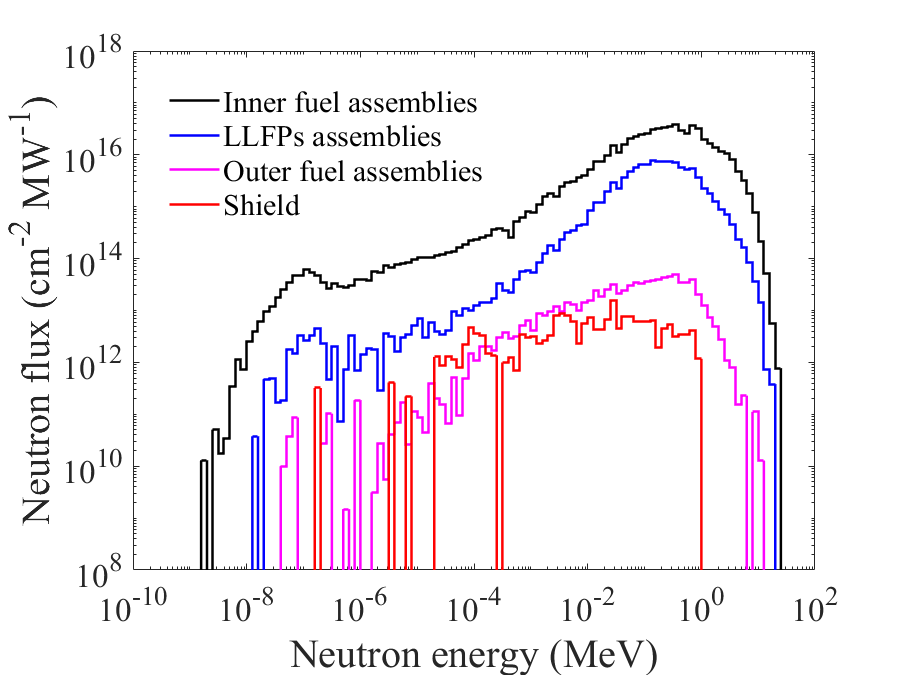}
    \caption{Spectral patterns of neutrons diagnosed at different layers of ANES with the $^{235}$U enrichment in the fuel assembly of $25.25\%$.}
    \label{fig6}
\end{figure}

The neutron spectra in different regions of the subcritical reactor core are shown in Fig.~\ref{fig6}. The neutron flux decreases along the radial direction, in particular in the  LLFPs assembly, where neutrons are absorbed by the transmutation processes. In the region of LLFPs assembly, the shape of the neutron spectrum is very similar to those in the region of the inner fuel assembly. 
In the shield region, the neutron flux is three orders of magnitudes lower than that in the inner region of the fuel assembly, indicating  effective shielding that reduces significantly the neutron radiation coming from ANES.

\begin{table*}[ht]
\centering
\small\addtolength{\tabcolsep}{8.pt}
\renewcommand{\arraystretch}{1.1}
\caption{The key parameters of ANES in the initial moment.}
\begin{tabular}{lr}\hline
Physical quantity                                        & Value  \\ \hline
Effective multiplication factor $k_{\rm eff}$                   & 0.979  \\
Reactivity $\rho$                                           & -0.021 \\
Effective multiplication factor for prompt neutrons $k_{p}$ & 0.978  \\
Eigenvalue $\alpha$                                           & -0.003 \\
Effective delayed neutron fraction $\beta_{\rm eff}$                & 0.007  \\
Neutron generation time $\Lambda$ ($\mu$s)                         & 0.523  \\
Neutron worth of PNS $\varphi$                                & 3.07\\
Subcritical effective multiplication factor $k_{s}$         & 0.993   \\ \hline
\end{tabular}
\label{Tab2}
\end{table*}

The performance of ANES can be quantified in terms of:  $k_{\rm eff}$, $k_{s}$, the reactivity $\rho$, 
the effective multiplication factor for prompt neutrons $k_{p}$, 
the effective delayed neutron fraction $\beta_{\rm eff}$, the neutron generation time $\Lambda$ and the neutron worth $\varphi$ \cite{2011Calculation,verboomen2006monte}. Values of these parameters for the GF-beam-driven ANES are collected in Table~\ref{Tab2}. The initial value of $k_{\rm eff}$ is $0.979$. During two years of burn-up, $k_{\rm eff}$ decreases to $0.902$. 
The system reactivity $\rho\,$$=$$\,(k_{\rm eff}-1)/k_{\rm eff}$ is then calculated to be less than $-0.021$. Such a negative $\rho$ value measures the safety of the proposed ANES setup.

\begin{figure*}[htbp]
\centering
    \includegraphics[width=.9\textwidth]{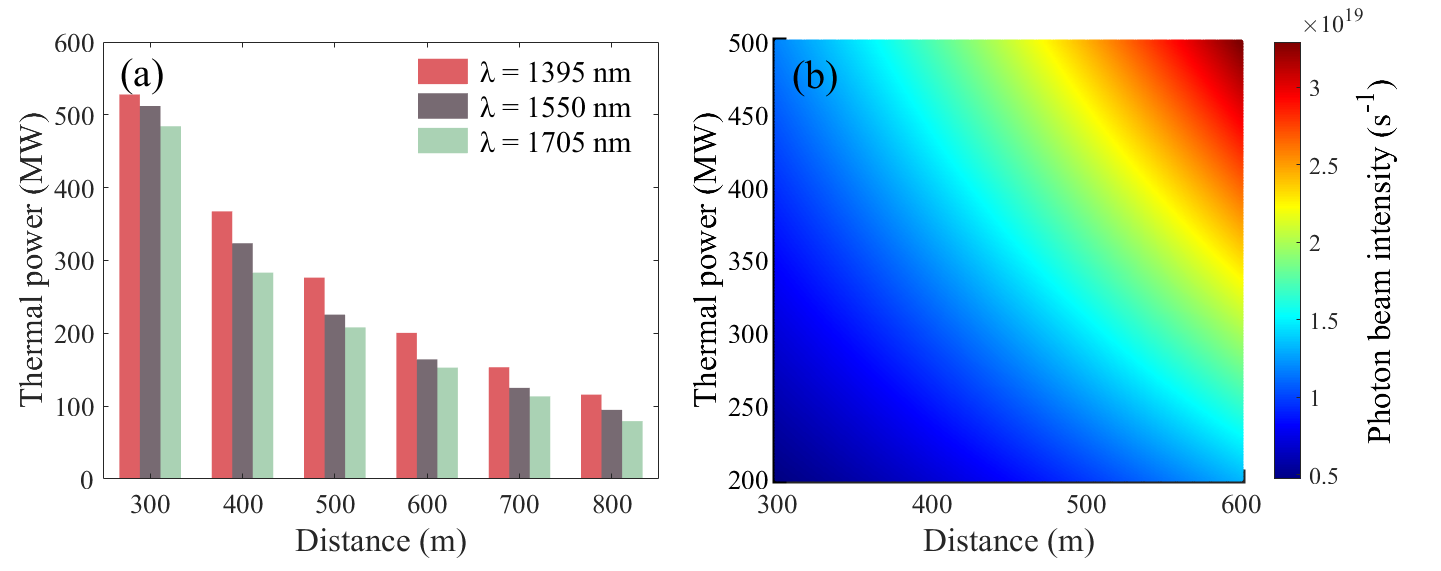}
    \caption{The thermal power $P_{t \gamma}$ of the subcritical reactor as a function of the distance $d$ between the GF photon source 
    and ANES for the spectrum-integrated photon intensity $I_\gamma\,$$=$$\,10^{19}\,\gamma/$s at IP (a),
    and the dependence of the required $I_\gamma$ on the thermal power $P_{t \gamma}$ and the distance $d$ (b). 
    In panel (b), the values of the key parameters used in the calculations are: $P_{\gamma n}\,$$=$$\,0.023$, 
    $\varphi\,$$=$$\,3.07$ and $k_{\rm eff}\,$$=$$\,0.979$.}
    \label{fig7}
\end{figure*}

The ANES thermal power $P_{t \gamma }$ can be evaluated%
\footnote{The operation-time dependence of $P_{t \gamma }$ is not taken into account. The reactor thermal power should be considered as the initial one at the starting point of the ANES operation. It will decrease with the speed of burning of nuclear fuel unless the GF photon flux is increased to compensate for the diminishing flux of fission neutrons.} 
by the following formula \cite{N22sun2022transmutation}:
\begin{equation}
P_{t \gamma } =E_{f} \cdot{I_{\gamma }} \cdot{P_{\gamma n}} \cdot{\frac{k_{\rm eff}}{1-k_{\rm eff}} }
\cdot{\frac{1}{\bar{v} }}\cdot{\varphi}\,, 
\label{eq:Ptgam}
\end{equation}
where $I_{\gamma}$ is the $\gamma$-ray beam intensity at IP and $\bar{v}$ is the average number of fission neutrons. The dependence of $P_{t \gamma}$ on $d$ is shown in Fig.~\ref{fig7}(a). At  $\lambda\,$$=$$\,1395\,$nm, $P_{t \gamma }$ reaches the value of $528\,$MW. It is slightly higher than those obtained for the two other laser wavelengths owing to the highest photo-neutron yield, as shown in Fig.~\ref{fig3}(a). 

According to Eq. (\ref{eq:Ptgam}), the spectrum-integrated photon-beam intensity $I_{\gamma }$ which is necessary to reach the required $P_{t \gamma }$ depends upon the distance $d$ between the ANES and photon-source positions. A contour plot for such dependence is shown in Fig.~\ref{fig7}(b). It indicates that reaching higher $P_{t \gamma }$ requires larger $I_{\gamma }$ or the smallest allowed distance $d$.
If ANES is positioned $320\,$m away from the GF photon source, its initial thermal power exceeds $500\,$MWt when 
$I_{\gamma}\,$$=$$\,10^{19}\,\gamma/$s. 


To transmute LLFPs efficiently, the minimal thermal power of the order of $100\,$MWt is required. For such a power, $I_{\gamma}$ in the GF-driven ANES should exceed $10^{18}\,\gamma/$s, which is $\sim\,$$5.5$ times lower than the one estimated in \cite{N22sun2022transmutation} for an idealized and unrealistic case of ANES placed right at the position of the electron-beam-driven photon source.

\section{Transmutation of LLFPs}
\label{sec:transmutation}

\begin{figure}[htbp]
\centering
    \includegraphics[width=.5\textwidth]{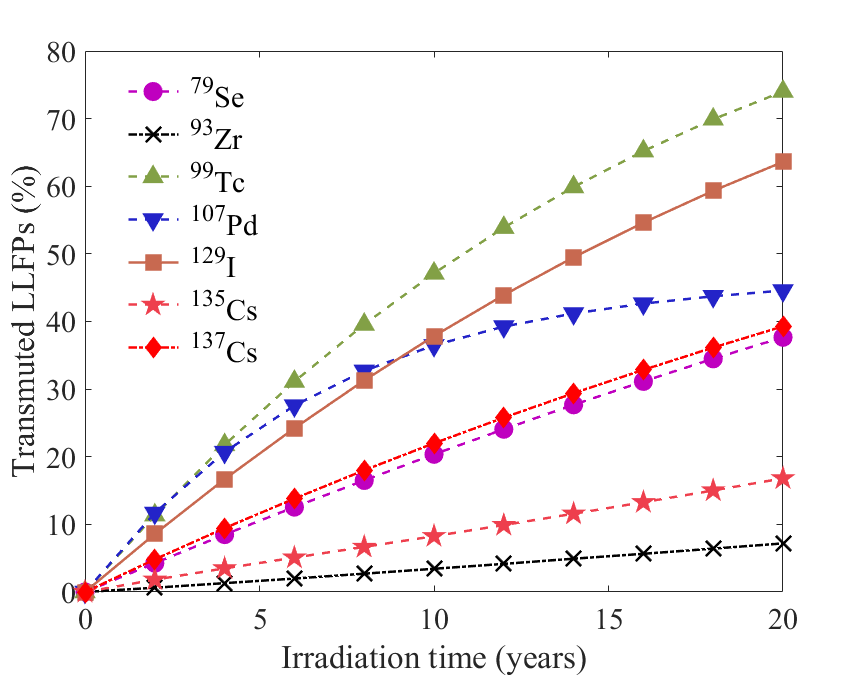}
    \caption{Transmutation of LLFPs over $20$ years irradiation for the initial thermal reactor power of $500\,$MWt. The percentages of transmuted LLFPs after $20$ years of the ANES operation are in the following order: $^{99}$Tc$\,\leq\,$$^{129}$I$\,\leq\,$$^{107}$Pd$\,\leq\,$$^{79}$Se$\, \approx\, $$^{137}$Cs$\,\leq\,$$^{135}$Cs$\,\leq\,$$^{93}$Zr.}
    \label{fig8}
\end{figure}

The time dependence of transmuted LLFPs over $20$ years of continuous operation of the GF-beam-driven ANES  has been simulated. 
As shown in Fig.~\ref{fig8}, the mass of the transmuted LLFPs in the LLFPs assembly increases approximately linearly with 
the ANES operation time. During the $20$-year-long operation time, the reduction of the 
number of the $^{79}$Se, $^{99}$Tc, $^{107}$Pd, $^{129}$I and $^{137}$Cs isotopes is higher than $35\%$, 
whereas for $^{93}$Zr and $^{135}$Cs, the transmutation rates are less than $20\%$. 
Similar results have been obtained in the fast neutron transmutation design \cite{N32chiba2017method}. 
The low transmutation efficiency for $^{93}$Zr and $^{135}$Cs isotopes is mainly due to their relatively small neutron capture cross sections.

\begin{table*}[ht]
\small\addtolength{\tabcolsep}{7.pt}
\renewcommand{\arraystretch}{1.5}
\centering
\caption{The values of the LLFPs parameters obtained from the {\sf MCNPX} simulations considering LLFPs transmuted at the thermal power of $500\,$MWt.}
\begin{tabular}{lrrrrrrr}\hline
LLFPs                        & $^{79}$Se & $^{93}$Zr  & $^{99}$Tc & $^{107}$Pd & $^{129}$I & $^{135}$Cs & $^{137}$Cs \\ \hline
$T_{\rm eff}$  (year)            & 26.6 & 135.4 & 13.5 & 22.5  & 15.7 & 59.5  & 25.5  \\
TR (\%/year)                 & 1.88 & 0.37  & 3.7  & 2.22  & 3.18 & 0.84  & 1.96  \\
\makecell[l]{Transmutation in LLFPs\\ assembly (g/year)} & 53   & 795   & 7980 & 473   & 1530 & 2860  & 6840  \\
\makecell[l]{Production in fuel\\ assembly (g/year)}     & 51   & 4330  & 4340 & 396   & 824  & 6900  & 6500  \\
SR                           & 1.04 & 0.18  & 1.84 & 1.19  & 1.86 & 0.41  & 1.05  \\ \hline
\end{tabular}
\label{Tab3}
\end{table*}

The rate of transmuted LLFPs is used to evaluate the effective half-life $T_{\rm eff}$ \cite{N22sun2022transmutation}, the transmutation rate TR and the support ratio SR for these LLFPs.
$T_{\rm eff}$ is defined as the effective half-life of LLFPs considering both the transmutation process and the natural decay in the reactor core. TR is the ratio of the mass of the transmuted LLFPs to that of the initially loaded ones,  and SR is the ratio of the mass of the transmuted LLFPs to that of the produced ones. 
The values of these parameters for the GF-driven ANES are shown in Table~\ref{Tab3}. $T_{\rm eff}$ of the transmuted LLFPs in the GF beam-driven ANES is $\sim\,$$100$ years. This value is significantly smaller than the one for non-transmuted LLFPs for which 
$T_{\rm eff}$ is larger than $10^6$ years. 
For the $^{99}$Tc, $^{107}$Pd, $^{129}$I and $^{137}$Cs isotopes, 
TR reaches $\sim\,$$2\,$--$\,3\%$ per year. 
SR is larger than $1.0$ for $^{79}$Se, $^{99}$Tc, $^{107}$Pd, $^{129}$I and $^{137}$Cs, indicating their decreasing yields as a function of the operation time of ANES. For the $^{93}$Zr and $^{135}$Cs isotopes, $SR$ is smaller than $1.0$ both due to the small neutron capture cross sections and the large production yields. It should be noted that the averaged TR over the seven LLFPs can reach $2.02\%$ per year for the GF-beam-driven ANES. This is slightly higher than $1.94\%$ per year given in \cite{N22sun2022transmutation} and $1.51\%$ per year in a fast reactor system \cite{N32chiba2017method}. 

\section{Discussion}
\label{sec:discussion}

\begin{figure}[htbp]
\includegraphics[width=0.5\textwidth]{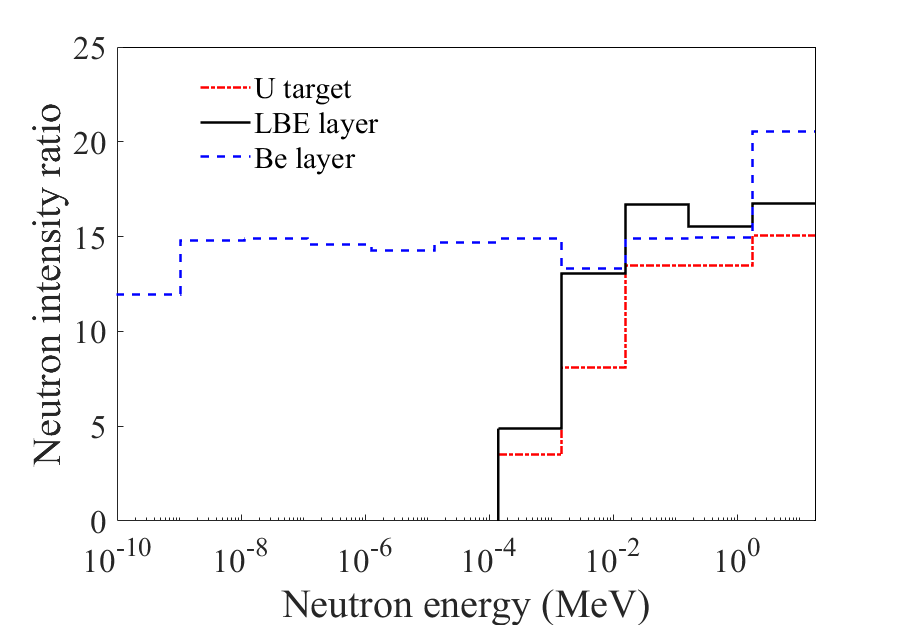}
\caption{The intensity ratio of neutrons escaped from different layers of the proton-based ANES to that of the GF-based ANES (per MW of the beam power). The energy of the proton beam used in the {\sf Geant4} simulations is $500\,$MeV, and the thicknesses of the LBE and Be layers are $T_{\rm LBE}\,$$=$$\,2\,$cm and $T_{\rm Be}\,$$=$$\,16\,$cm, respectively.}	
\label{fig9}
\end{figure}

In this section, we discuss the relative advantages and disadvantages of the photon-beam-based ANES and the proton-beam-based one. Additional {\sf Geant4} simulations have been performed to study the neutron production in metallic $^{238}$U irradiated by the $500\,$MeV proton beam. Note that such a proton beam could be produced by the China initiative Accelerator Driven System \cite{N11gu2021latest}. The simulation results show that each proton produces on average $\sim\,$$18$ neutrons. These proton-neutrons have the average energy of $12.32\,$MeV, which is significantly higher than $5.96\,$MeV obtained with the GF $\gamma$-ray beam. The $1\,$MW proton beam can then produce  
$\sim\,$$2.25\times 10^{17}$ neutrons escaping from the PNS target. The Be-layer moderated neutron flux generated by the proton beam (per MW of the beam power) is thus approximately $14$ times higher than that of the GF $\gamma$-ray beam of the same power, as shown in Fig.~\ref{fig9}.

\begin{figure}[htbp]
\includegraphics[width=0.5\textwidth]{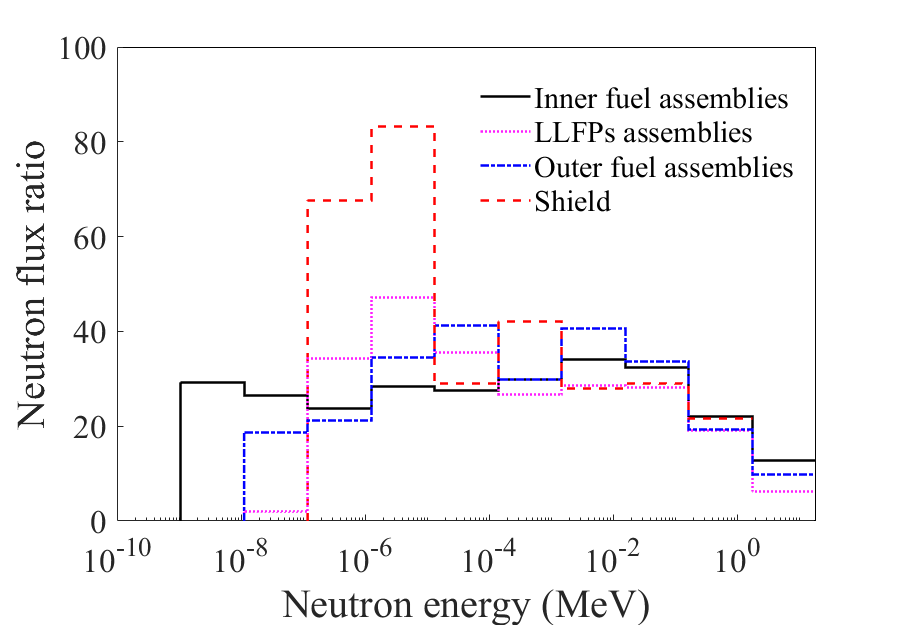}
\caption{The ratios of the neutron fluxes diagnosed at different regions of the proton-based ANES to that of the GF-based ANES (per MW of the beam power). The $^{235}$U enrichment is $25.25\%$.}	
\label{fig10}
\end{figure}

The ratios of the neutron flux in different regions of ANES for the proton and photon beams  of the same beam power are shown in Fig.~\ref{fig10}. The proton beam produces over $20$ times higher the neutron flux in the reactor core compared to the photon-beam-driven scheme. This is attributed to a higher neutron production efficiency and a larger proton-neutron worth $\varphi_{p}$ $\sim\,$$4.2$, compared to $\varphi\,$$=$$\,3.07$ for the photon-beam case. The neutron spectra of the proton-beam-based ANES are similar to those of the photon-beam one. The neutron flux in the proton-driven ANES decreases along the radial direction in a similar way as for the photon-driven one. 

The thermal power of the proton-beam-driven ANES can be expressed in a similar way as for the photon-beam-driven one:
\begin{equation}
P_{t p } =E_{f} \cdot{I_{p } } \cdot{P_{p n}} \cdot{\frac{k_{\rm eff}}{1-k_{\rm eff}} }\cdot{\frac{1}{\bar{v} }}\cdot{\varphi_{p}}\,,
\label{eq:Ptp}
\end{equation} 
where $I_{p}$ is the proton beam intensity and $P_{p n}\,$$=$$\,18.4$ is 
the neutron production rate. According to Eqs.~(\ref{eq:Ptgam}) and (\ref{eq:Ptp}), the photon and proton beam powers required to drive the thermal power of $500\,$MWt are evaluated to be $\sim\,$$16.4$ and $0.86\,$MW, respectively. 

Compared to the GF $\gamma$-ray beam, the operation  of a high-energy proton beam of the same MW power requires higher electrical power of an accelerator site. For example, to generate the proton beam of $0.75\,$MW power, the Japan Proton Accelerator Research Complex (JPARC) requires the electrical power of $60\,$MW \cite{JPARC2018capacitor}. To produce the $\gamma$-ray beam power of $16.65\,$MW, GF would need the electrical power of $\sim\,$$50\,$MW%
\footnote{The main electric power consumers for the GF photon beam production scheme are: (1) the LHC cryogenic system  and (2) the LHC RF system. Note that the rigidity of the PSI beam at the LHC is sufficiently high to allow for multiple emissions of the $\gamma$-rays over the PSI revolution time, with negligible particle-beam losses. In the GF scheme, the power of the RF system can thus be directly converted,  by the stored PSI beam,  into the power of the GF photon beam.}. 
The above comparison shows that the electrical power required to produce a $1\,$MW proton beam is $26.5$ times higher than that of GF $\gamma$-ray beam. Therefore, even if the $\gamma$-ray beam has the lower neutron production efficiency, the overall electrical power efficiency of the GF-driven ANES could be comparable to that of the proton-beam-driven ANES.  It remains to be added that while producing the MW-class proton beams represents a serious technological challenge, related to the beam-dynamics effects that are specific to high-density bunches of charged fermions, there is no intensity limit for the high-density ``bosonic'' photon beams. 
Its power is determined by the  RF power of the PSI storage ring. As a fraction of the ANES power can be used to generate the requisite RF power, the photon-beam-driven scheme may become more attractive for high-power ANES facilities.   

\begin{table*}[ht]
\centering
\small\addtolength{\tabcolsep}{5.pt}
\renewcommand{\arraystretch}{1.5}
\centering
\caption{The evaluated masses of LLFPs transmuted by a direct photo-nuclear reaction using the GF $\gamma$-ray beam with $\lambda\,$$=$$\,1550\,$nm. $E_{(\gamma,n)}$ and $E_{(\gamma,2n)}$ represent the threshold energies for the $(\gamma,n)$ and $(\gamma,2n)$ reactions, respectively.}
\begin{tabular}{lrrrrrrr} \hline
LLFPs                 & $^{79}$Se &$^{93}$Zr  & $^{99}$Tc &$^{107}$Pd &$^{129}$I &$^{135}$Cs &$^{137}$Cs \\ \hline
Proportion of isotopes (\%)                      & 12.12 & 20.33  & 100.00 & 18.3   & 23.98  & 31.87  & 32.59  \\
$E_{(\gamma,n)}$ (MeV)                             & 7.00  & 7.00   & 8.99   & 7.00   & 8.87   & 8.77   & 8.90   \\
$E_{(\gamma,2n)}$  (MeV)                           & 17.50 & 15.50  & 16.50  & 16.50  & 16.00  & 16.00  & 15.50  \\
\makecell[l]{Transmutation in LLFPs\\ target (g/year)}& 5.69  & 20.94  & 97.54  & 21.31  & 20.29  & 32.56  & 33.04  \\ \hline
\end{tabular}
\label{Tab4}
\end{table*}

There is also another reason to develop the photon-beam-driven ANES. The PNS-driven transmutation scheme which is common to both the proton and photon-driven ANES can be complemented, only in the latter case, by the direct photon transmutation \cite{N33hayakawa2016proposal}.  
The GF photon beam covers the energy domain of $7\,$--$\,9\,$MeV. Photons in this energy domain can transmute, via resonant photo-fission
processes, those of LLFPs which are abundantly produced in the fuel assembly and remain resistant to neutron transmutation \cite{N22sun2022transmutation}. 

The efficiency of the photo-transmutation induced by the GF $\gamma$-rays in this energy range is estimated by assuming the following composition of the photon-beam target: selenium, iodine and cesium are loaded in form of compounds, like ZnSe and CsI, to ensure stable
chemical forms, while zirconium, technetium and palladium are loaded in metallic forms \cite{N22sun2022transmutation}. 
Such a target would have to be placed in front of ANES and should cover the angular range allowing to capture the $7\,$--$\,9\,$MeV photons.
Table~\ref{Tab4} shows the isotopic proportions, the photo-neutron reaction thresholds and the annual transmutation masses of these LLFPs. One can see that the annual transmutation masses of $^{99}$Tc, $^{129}$I and $^{137}$Cs are an order of magnitude lower than those in ANES. However, for the $^{93}$Zr and $^{135}$Cs isotopes for which the neutron-transmutation SR is smaller than $1.0$, as shown in  Table~\ref{Tab3},  the photo-fission would be the only way to partially transmute $21$ and $32\,$g/year of these isotopes. Moreover, using the optimized $\gamma$-ray beam at energies between $7$ and $9\,$MeV for the photo-transmutation of $^{93}$Zr and $^{135}$Cs will offer a unique opportunity for the safe disposal of these LLFPs
-- something that is not reachable for the proton-based ANES.

\section{Conclusions}

In this paper, we have presented the exploratory studies of the GF photon-beam-driven ANES, generating up to $500\,$MW of 
the thermal power with the efficient transmutation of the loaded and produced LLFPs.  
Such an ANES is driven by the high-flux PNS which is generated by the optimized GF $\gamma$-ray beam. 
The performance of the GF-based ANES and the resulting transmutation capability have been analyzed. 
For the GF-driven ANES operating at the thermal power of $500\,$MW over $20$ years, the ratios of the amount of the transmuted LLFPs to that of the produced ones, SRs,  are larger than $1.0$ for the $^{79}$Se, $^{99}$Tc, $^{107}$Pd, $^{129}$I and $^{137}$Cs isotopes. 
The effective half-lives of these LLFPs, $T_{\rm eff}$, can be reduced from almost $10^6$ years to about $100$ years, 
which dramatically decreases their cooling times. 

Compared to the proton-based ANES, the relative advantages and disadvantages of the GF-based ANES have also been evaluated. 
Although the neutron production efficiency of the GF $\gamma$-ray beam (per MW of the beam power) is about $14$ times lower 
than that of the $\sim\,$$1\,$GeV proton beam, the GF photon-beam-driven ANES may have a comparable net electrical power supply efficiency 
and could provide a complementary transmutation method for those of LLFPs that are resistant to the neutron-driven transmutation.

\appendix 

\section{Methods and materials}
\label{sec:metmat}

\subsection{Computational models and methods}
\label{subsec:comp}

The GF $\gamma$-ray source generation was simulated with {\sf GF-CAIN} \cite{GF-CAIN}
which is a GF-customized \cite{GF-PoP-LoI:2019,Bieron:2021ojp,Apyan:2022ysh} version of the Monte Carlo program {\sf CAIN} \cite{CAIN} 
developed at KEK--Tsukuba, Japan, for the ILC project \cite{ILC}.
The beam parameters of the PSI and Erbium laser shown in Table~\ref{tab:YbHepars} were employed as input for these simulations. 

Then, the GF $\gamma$-ray source was imported into {\sf Geant4} (version {\sf 4.10.3}) \cite{2018The}, which was used to simulate the PNS generation and the following neutron moderating processes. Considering that the photo-neutrons result mainly from the photo-neutron and photo-fission reactions on the $^{238}$U target, we then employed the {\sf QGSP} model in {\sf Geant4} to simulate the former reaction and {\sf Geant4-GFDPO} \cite{N27shi2024geant4} to model the latter. 

The production of the proton-neutrons was also simulated with {\sf Geant4} by invoking the {\sf BERT\_HP} model \cite{1968Calculation,1971Calculation}. Accordingly, the energy, position and momentum distributions of both PNS and the proton-neutron source were obtained. This source information was further imported into {\sf MCNPX} (version {\sf 2.7}) \cite{Pelowitz2021MCNPX} which was used to evaluate the performance of ANES. The reaction cross sections required for the {\sf MCNPX} simulations were taken from the {\sf ENDF-VII.1} library \cite{database2021}. In the {\sf MCNPX} simulations, the necessary physical processes, including the neutron-capture, neutron-fission and photo-nuclear reactions were taken into account. In addition, the burn-up calculations of the GF-based ANES were performed using {\sf MCNPX}. The effective neutron multiplication factor $k_{\rm eff}$ has a statistical error lower than $0.1\%$ and the reaction rate for evaluation of the transmutation efficiency is within the $0.5\%$ statistical precision.

\subsection{Selection of LLFPs}
\label{subsec:LLFPs}

The major selected LLFPs are $^{79}$Se, $^{93}$Zr, $^{99}$Tc, $^{107}$Pd, $^{129}$I, $^{135}$Cs and $^{137}$Cs \cite{N37kailas2015nuclear,N38yang2004long,N39wang2020spallation}. These radionuclides can cause long-term radioactivity during the geological disposal of SNF \cite{N19optimization2017}. The compositions of LLFPs were obtained from the burn-up simulation of uranium dioxide pellets by fast breeder reactor core at $50\,$GWd/t for two years. Without the isotopic separation, such compositions were used as the initial compositions of LLFPs in the pins. The details of these compositions are presented in \cite{N22sun2022transmutation}.

\section{Selection of $^{238}$U target}
\label{sec:Utarget}

The metallic $^{238}$U was selected for photo-neutron production, since it is widely used in nuclear energy, easy to obtain without enrichment process and has a large neutron production cross section \cite{N40filipescu2023photofission}. Particularly, $^{238}$U target possesses an excellent neutron generation property in the case of small geometry with a radius less than $15\,$cm \cite{N41ismailov2011feasibility}. The LBE coolant temperature of the lead-based fast reactor ranges from $400$ to \SI{600}{\degreeCelsius} \cite{N42loewen2003status}, which is visibly lower than the melting point of the $^{238}$U target ($\sim\,$\SI{1132}{\degreeCelsius}). In our design, the $^{238}$U target is optimized with a dual-layer structure which possesses a channel in the middle for the flow of LBE, as shown in Fig.~\ref{fig1}(b). The entire target is immersed in the coolant to ensure effective heat dissipation and prevent the target from melting.

\subsection*{Acknowledgments}
This work has been supported by the National Key R\&D Program of China (Grant No. 2022YFA1603300) 
and the National Natural Science Foundation of China (Grant No. U2230133). 
The research of WP has been supported in part by a grant from the Priority Research Area (DigiWorld) under the Strategic Programme Excellence Initiative at the Jagiellonian University in Krakow, Poland.

\bmhead{Data Availability Statement}
No data is associated with the manuscript.



\end{document}